\documentclass[aps,prb,superscriptaddress,notitlepage,nopacs,amsmath,amstex,amssymb,citeautoscript,longbibliography,floatfix,letter,twocolumn]{revtex4-2}
\usepackage{graphicx}
\usepackage{adjustbox}
\usepackage{bm}
\usepackage{ulem}
\usepackage{color}
\usepackage{braket}
\usepackage{standalone}
\usepackage{multirow}
\usepackage{tikz}
\usepackage{mathrsfs}
\usepackage{dsfont}
\usepackage{comment}
\usepackage{amsmath, amsthm}

\usepackage[colorlinks,bookmarks=true,citecolor=blue,linkcolor=red,urlcolor=blue]{hyperref}
\usepackage{cleveref}
\usepackage{orcidlink}

\graphicspath{{./figures/}}
\begin{document}

\title{Monte Carlo Sampling for Wavefunctions Requiring Symmetrization or Antisymmetrization}

\author{Koyena Bose}
\email{koyenab@imsc.res.in}
\affiliation{Institute of Mathematical Sciences, CIT Campus, Chennai 600113, India}
\affiliation{Homi Bhabha National Institute, Training School Complex, Anushaktinagar, Mumbai 400094, India} 

\author{Steven H. Simon}
\email{steven.simon@physics.ox.ac.uk}
\affiliation{Rudolf Peierls Centre, Oxford University, OX1 3NP, United Kingdom}

\author{Ajit C. Balram\orcidlink{0000-0002-8087-6015}}
\email{cb.ajit@gmail.com}
\affiliation{Institute of Mathematical Sciences, CIT Campus, Chennai 600113, India}
\affiliation{Homi Bhabha National Institute, Training School Complex, Anushaktinagar, Mumbai 400094, India}

\date{\today}

\begin{abstract}
Many strongly correlated states, such as those arising in the fractional quantum Hall effect and spin liquids, are described by wavefunctions obtained by dividing particles into multiple clusters, constructing a readily evaluable wavefunction in each cluster, and (anti)symmetrizing across these clusters. We introduce a method to compute quantities such as energies and correlators, using Monte Carlo simulations for these states. Our framework overcomes the factorial scaling of explicit (anti)symmetrization, allowing for studies of systems beyond the reach of exact diagonalization.
\end{abstract}

\maketitle

\textbf{\textit{Introduction.}} Strongly interacting systems can exhibit phases characterized by topological order that lies beyond the conventional Landau paradigm of symmetry-breaking of a local order parameter. These topological phases are characterized by long-range entanglement~\cite{Wen13}, ground-state degeneracy on nontrivial manifolds~\cite{Wen90d}, and braiding of the anyonic quasiparticles they host~\cite{Arovas84, Nayak08}. Experimentally, these phases are realized in the fractional quantum Hall effect (FQHE)~\cite{Tsui82}, where electrons confined to two dimensions (2D) are subjected to a strong perpendicular magnetic field. In this regime, the single-particle spectrum organizes into discrete Landau levels (LLs), and within a partially filled LL, kinetic energy gets quenched, and interactions dominate. At selected rational fillings of an LL, interactions can open up a gap, resulting in incompressible FQHE phases. These states exhibit many interesting properties, foremost among which are the existence of fractionally charged excitations that exhibit anyonic statistics~\cite{Wilczek82, Laughlin83, de-Picciotto97, Nakamura20, Bartolomei20}. In certain FQHE states, these excitations exhibit non-Abelian braiding statistics that could, in principle, enable building a fault-tolerant topological quantum computer~\cite{Freedman03, Nayak08}.

A prominently observed even-denominator state occurs at filling $\nu{=}5{/}2$~\cite{Willett87}, which is widely believed to be non-Abelian~\cite{Banerjee17b, Dutta21} since one of the leading candidates to describe this FQHE is the Moore–Read (MR) Pfaffian (Pf)~\cite{Moore91} state, or its hole-conjugate the anti-Pf~\cite{Levin07, Lee07, Rezayi17}. The MR is the $k{=}2$ member of a broader class of $k$-clustered Read-Rezayi (RR$k$) states, which are non-Abelian for $k{>}1$. Interestingly, these states can be written as conformal field theory (CFT) correlators of parafermionic primary fields, allowing for a direct study of their non-Abelian properties via conformal blocks~\cite{Zamolodchikov85, Moore91, Nayak96, Read99}. Members of the RR$k$ series could underpin some experimentally observed fractions in the second LL (SLL) (and its analog in bilayer graphene~\cite{Zibrov16, Huang21, Assouline23, Kumar24, Huang25}), in particular, the $12/5$ state~\cite{Xia04, Pan08, Choi08, Kumar10, Zhang12} has been proposed to be described by the hole-conjugate of the RR$3$ state~\cite{Read99, Wojs09, Mong15, Zhu15, Pakrouski16, Balram21b}. Furthermore, bosons in the lowest LL (LLL) interacting via the ultra-short-range delta-function or the long-range Coulomb interaction can stabilize the MR and the RR$3$ states~\cite{Cooper01, Regnault03, Regnault04, Sharma23}. 

The bosonic RR$k$ wavefunction is obtained by symmetrizing over partitions of $N$ particles into $k$ clusters ($N$ and $k{>}1$ are positive integers with $N$ divisible by $k$), where each cluster forms a bosonic $1{/}2$ Laughlin state~\cite{Cappelli01}. Being a combinatorial operation, numerical construction of RR$k$ wavefunctions by explicit symmetrization is computationally prohibitive except for very small systems. The RR$k$ states can be alternatively constructed via exact diagonalization (ED)~\cite{Read99} and Jack polynomials~\cite{Bernevig08, Bernevig08a}, but these procedures are likewise limited to small sizes ($N{\leq}25$). The RR$3$ state can also be generated using matrix product states on the quasi-$2$D cylindrical geometry for fairly large systems~\cite{Wu14b, Estienne15}, but not on the sphere~\cite{Haldane83}, torus~\cite{Haldane85}, or disk~\cite{Laughlin81}, which are the conventional $2$D geometries for FQHE studies. Readily evaluable wavefunctions that lie in the same universality class as the hole-conjugate of the fermionic RR$k$ states exist~\cite{Balram19}, but they lack the desirable features of RR$k$ states, such as the existence of a model Hamiltonian or being exact CFT correlators. It is therefore useful to develop an approach that enables the evaluation of quantities of interest for the RR$k$ states for large systems in $2$D geometries. Here, we present such a method, allowing efficient computation of quantities such as energies and correlation functions for the RR$k$ states using Monte Carlo (MC). Our method is general and extendable to other wavefunctions that require explicit (anti)symmetrization, and could be applicable in other domains, such as to quantum spin liquids~\cite{Greiter11a}, for computing electronic structures~\cite{Bajdich06, Bajdich08}, describing molecules~\cite{Sorella07, Toulouse08, Zen14}, etc.

\textbf{\textit{Method.}} 
In this work, we demonstrate a way to do MC simulations of (anti)symmetrized wavefunctions that scales favorably with $N$, making it numerically tractable, unlike brute-force (anti)symmetrization. We use the $\nu{=}1$ bosonic MR state~\cite{Moore91} as an example to illustrate our idea. We write its wavefunction as~\cite{Ho95}
\begin{equation}
\label{eq: bosonic_MR_1}
    \Psi^{\rm MR}_{1}{=} \mathbb{S}\left[ \Phi^{2}_{1}(z_{1},{\cdots},z_{N/2})~\Phi^{2}_{1}(z_{N/2{+}1},{\cdots},z_{N}) \right],
\end{equation}
where $\Phi_{1}(z_{1},{\cdots},z_{M}){=}\prod_{1{\leq}i{<}j{\leq}M}(z_{i}{-}z_{j})$ is the $\nu{=}1$ integer quantum Hall state. The wavefunction of Eq.~\eqref{eq: bosonic_MR_1} is obtained by dividing the particles into two clusters, forming the bosonic $\nu{=}1/2$ Laughlin state~\cite{Laughlin83} in each of the clusters, i.e., the Halperin (H)-$(2,2,0)$ state~\cite{Halperin83}, 
\begin{equation}
\label{eq: bosonic_Halperin_220}
    \Psi^{{\rm H}(2,2,0)}_{1}{=} \left[ \Phi^{2}_{1}(z_{1},{\cdots},z_{N/2})~\Phi^{2}_{1}(z_{N/2{+}1},{\cdots},z_{N}) \right],
\end{equation}
and symmetrizing over all the particles, which the operator $\mathbb{S}$ implements. Here, one could write a Pf for MR, which can be evaluated efficiently. However, assuming the Pf form were not known and one instead performs the explicit symmetrization, the computational complexity would scale poorly with increasing $N$. By contrast, it is easier to compute a matrix element like $\langle \Psi^{\rm MR}_{1} | \hat{O} | \Psi^{\rm MR}_{1} \rangle$, where $\hat{O}$ is a symmetric operator. Specifically, expectation values of interactions of the form $\hat{V}{=}\sum_{i{<}j}V_{i,j}$ such as the Coulomb interaction, where $V_{i,j}{=}1/|z_{i}{-}z_{j}|$, are simpler to evaluate than the wavefunction itself since these interactions are invariant under particle permutations. To demonstrate this, we write the symmetrization as a sum over permutations,
\begin{equation}
   \Psi^{\rm MR}_{1}{=}\frac{1}{N!}\sum_{P{\in}S_{N}} P\Psi^{{\rm H}(2,2,0)}_{1},
\end{equation}
where $S_{N}$ is the symmetric group, and $P g(z_{1},{\cdots}z_{N}){=}g(z_{P(1)},{\cdots},z_{P(N)})$ [Note that the total number of permutations for the $k{=}2$ case is fewer and equals $(N!)/(2! [(N/2)!]^{2})$, since permutations of indices within a cluster are identical and the two clusters can be exchanged overall.]. Our matrix element is then
\begin{eqnarray}
\label{eq: MR_matrix_element_sum_over_permutations}
    \langle \Psi^{\rm MR}_{1} | \hat{O} | \Psi^{\rm MR}_{1} \rangle&{=}&
    \left(\frac{1}{N!}\right)^{2} \sum_{P_{1}, P_{2}}  \langle \Psi^{{\rm H}(2,2,0)}_{1} | P_{1} \hat{O} P_{2} | \Psi^{{\rm H}(2,2,0)}_{1} \rangle  \nonumber \\
    &{=}& \frac{1}{N!} \sum_P  \langle \Psi^{{\rm H}(2,2,0)}_{1} | \hat{O} P | \Psi^{{\rm H}(2,2,0)}_{1} \rangle,
\end{eqnarray}
which follows because $P_{1} \hat{O} P_{2}{=} \hat{O} P_{1} P_{2}$, as $\hat{O}$ is invariant under permutations, i.e., particle-index exchanges. Hence, the double sum over $P_{1}$ and $P_{2}$ reduces to a single sum over all permutations $P$ for a fixed base choice of $\Psi_1^{H(2,2,0)}$[Here, we choose $(1,2,{\cdots}, N/2)$ and $(N/2{+}1,{\cdots}, N)$ as the base set of the two clusters.]. For a fixed base set, evaluating $\langle \Psi^{{\rm H}(2,2,0)}_{1} | \hat{O} P | \Psi^{{\rm H}(2,2,0)}_{1} \rangle$ yields only $\lfloor N/4 \rfloor{+}1$ distinct values, where $\lfloor x \rfloor{=}{\rm floor}(x)$ is the greatest integer${\leq}x$. This is because $\Phi^{2}_{1}$ is already fully symmetric in its arguments. Similarly, the two clusters can be swapped overall as they form the same state. Consequently, the matrix element $\langle \Psi^{{\rm H}(2,2,0)}_{1} | \hat{O} P | \Psi^{{\rm H}(2,2,0)}_{1} \rangle$ gives the same result for any $P$ which exchanges exactly $m$ of the indices ($1,{\cdots},N/2$) into the set ($N/2{+}1,{\cdots},N$), where $m{=}0, 1, {\cdots},\lfloor N/4 \rfloor$. Therefore, it suffices to evaluate the $\lfloor N/4 \rfloor{+}1$ distinct matrix elements by MC, and adding them with the appropriate combinatorial weights $C_{(N, m)}$ [see Supplemental Material (SM)~\cite{SM}] gives us the full matrix element $\langle \Psi^{\rm MR}_{1} | \hat{O} | \Psi^{\rm MR}_{1} \rangle$ that we desire. 

A subtle issue arises even when the operator $\hat{O}$ is well-behaved, i.e., is both Hermitian and invariant under the permutation of particles: the matrix elements we evaluate are generally complex (not necessarily real or positive), which can result in poor convergence. There are two reasons for this: (a) the sampling function is different from the wavefunction of interest, i.e., we use $H(2,2,0)$ for sampling instead of the fully symmetrized wavefunction; and (b) for each configuration corresponding to $m$ exchanges, which has $C_{(N,m)}$ terms in its ensemble~\cite{SM}, we evaluate only one representative term. While what we presented above is analytically correct, Metropolis MC using this approach is extremely slow to converge. This is because when we choose a particular representation, we have artificially chosen some particles to be different from others (those in the two different groups, and also those that are exchanged between groups). Thus, we must wait for the MC to sample configurations where all types of particles exchange positions with all other types of particles so that the average of the chosen representative faithfully emulates its ensemble's average. While eventually MC will sample the entire space equitably, there can be an extremely long so-called mixing time. 

In other words, if we include all terms in the full symmetrization, convergence to the expected value requires significantly fewer MC iterations. However, the cost per step becomes prohibitively expensive for large systems since the number of permutations scales factorially with $N$. On the other hand, approximating the many terms associated with each exchange by a single representative term is computationally inexpensive, but has a long mixing time and requires many MC iterations, which can again become time-consuming as the system size grows. Thus, the problem effectively becomes a trade-off between the number of terms to be included and the number of MC iterations required for convergence. For the naive method of evaluating exchanges, we use the form given in Eq.~\eqref{eq: MR_matrix_element_sum_over_permutations} and do MC sampling using the readily evaluable wavefunction $\Psi^{{\rm H}(2,2,0)}_{1}$. We compute the quantity $r{=}\sum_m \mathcal{R}_{N,m}$, where $\mathcal{R}_{N,m}{=}C_{(N,m)} r_m$, and $r_m{=}(\Psi^{{\rm H}(2,2,0),m}_{1} /\Psi^{{\rm H}(2,2,0)})$, where $\Psi^{{\rm H}(2,2,0),m}_{1}$ is a representative of the Halperin$-(2,2,0)$ state with $m$ exchanges between the two clusters. Using this, we compute the expectation value of the operator $\langle O \rangle$ as the ratio of the unnormalized symmetrized state [$\langle O\rangle_{\rm numerator}{=}\sum_{m} C_{(N,m)} \hat{O} r_m$], and the norm of the symmetrized state [by setting $\hat{O}{=}\mathbb{I}$, the identity operator, we have $\langle O\rangle_{\rm denominator}{=}\sum_{m} C_{(N,m)} r_m$].

In the refined version, we improve upon the naive method by strategically identifying and including additional terms to minimize error propagation. The key issue is that, even for a fixed $N$, the contribution or weight of each $m$-exchange is not uniform since $C_{(N,m)}$ increases with $m$. Thus, integrals for exchanges with smaller $C_{(N,m)}$ converge faster, since fewer correlations are encoded in them, than those with larger $C_{(N,m)}$. To mitigate this, we do a systematic correction by including more terms for larger $C_{(N,m)}$. This ensures that each $m^{\rm th}$ exchange contributes roughly uniform error to $\langle\hat{O}\rangle$, enabling convergence to the expected value within a reasonable time frame. Thus, we define the matrix element obtained by symmetrization via refined exchanges as
\begin{equation}
\label{eq: refined_exchanges}
    \langle \hat{O} \rangle_{\Psi^{\rm MR}_{1}}{=} \frac{1}{N!}  \langle \Psi^{{\rm H}(2,2,0)}_{1} | \hat{O}  \sum_m  \frac{C_{(N,m)}}{\mathcal{N}_{(N,m)}} \sum_{t=0}^{\mathcal{N}_{(N,m)}} |\Psi^{{\rm H}(2,2,0),m_{t}}_{1} \rangle,
\end{equation}
where $\mathcal{N}_{(N,m)}$ denotes the number of terms retained for the $m^{\rm th}$ exchange, and $t$ corresponds to a unique choice of $m$ exchanges between the two clusters. In other words, we group the large number of permutations into a much smaller number of equivalent groups. We then sample a few representative terms from each group and combine them with the correct combinatorial weights to obtain the matrix element.

Next, we explain how we choose $\{\mathcal{N}_{(N,m)}\}$ using the example of $N{=}20$. The values of $C_{(20,m)}$ are $1, 100, 2025, 14400, 44100, 31752$ for $m{=}0,1,2,3,4,5$, respectively. From a preliminary MC run, we observed that each exchange contributes roughly equally to the expectation $\langle\hat{O}\rangle$ for many observables $\hat{O}$. Since the combinatorial weight factor increases with increasing $m$, the numerator $\hat{O}r_m$, or $r_m$, decreases with increasing $m$. As a result, $r_5$ contributes the most error to $\langle O \rangle$, since its inaccuracy gets amplified when multiplied by its large combinatorial weight. To ensure that the error in each $r_m$ is of the same order, we define the number of terms to be computed for the $m^{\rm th}$ exchange [see Eq.~\eqref{eq: refined_exchanges}] as $\mathcal{N}_{(N,m)}{=}\lceil C_{(N,m)}{/}s_{\rm rep} \rceil$, where $\lceil x \rceil{=}{\rm ceil}(x)$, and $s_{\rm rep}$ is a parameter that is chosen to get to a desired accuracy. Based on the observation that $r_5$ has a value of the order of MC error, ${\sim}10^{-2}$, after $20$ million MC steps, which results in an error in the first significant digit, we choose $s_{\rm rep}{=}1000$ so that of the order $10$ additional terms are now included for $r_5$ which pushes the error in it down to the next significant digit. Accordingly, $\mathcal{N}_{(20,m)}{=}\lceil C_{(20,m)}{/}1000\rceil$, so that any $\mathcal{N}_{(20,m)}{>}1$ corresponds to including multiple representations (chosen arbitrarily) of the exchange of $m$ indices between the two clusters in the Halperin$-(2,2,0)$ state. As a result, instead of computing $\lfloor N/4 \rfloor{+}1{=}5$ ratios $r_{m}$, we compute a total of $\sum_m \mathcal{N}_{(N,m)}{=}97$ ratios, with $\mathcal{N}_{(N,m)}$ unique ratios for each $m$. This number is still much less than the minimal set of permutations, i.e., $20!/[(10!)^{2}2!]{=}92378$. 

\begin{figure}[htbp!]
\includegraphics[clip,width=0.49\columnwidth]{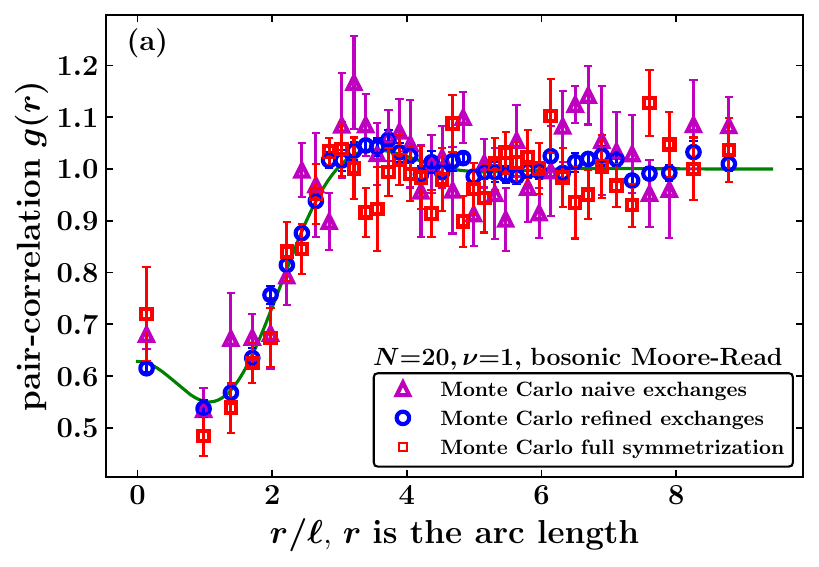}
\includegraphics[clip,width=0.49\columnwidth]{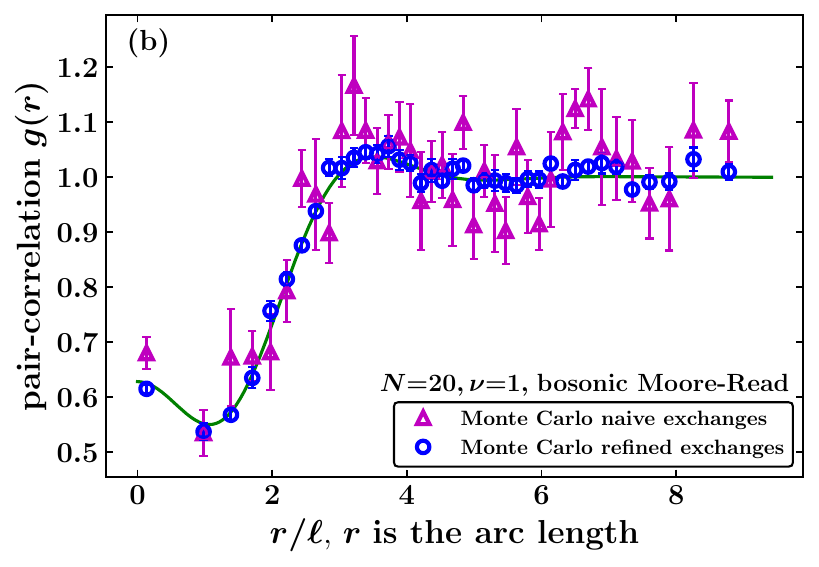}
\caption{(a) Pair-correlation function for the $\nu{=}1$ Moore-Read state of $N{=}20$ bosons on the sphere obtained with different methods keeping the same computational resources and runtime. (b) Same as (a), except, for clarity, results from only the refined and naive methods are shown.}
\label{fig: pair_correlation_bosons_N_20_comparison}
\end{figure}

Without refinement, the naive symmetrization has to be run for ${\sim} 100$ times more MC steps to achieve the same level of accuracy, i.e, to reduce the MC error to the next order. Note that since the naive method involves fewer terms per step and is therefore computationally faster than the refined method, it allows for more MC iterations in the same time. However, the refined method can be substantially accelerated through parallelization, yielding faster convergence with the same runtime, whereas multi-core processing offers limited benefits for the naive method. We explicitly show this in Fig.~\ref{fig: pair_correlation_bosons_N_20_comparison} by computing the pair-correlation function for $N{=}20$ using three methods: (i) naive, (ii) refined, and (iii) the full symmetrization. For each method, we run 10 MC chains, each for 10 minutes on a dual-Intel Xeon Platinum 8358@2.6 GHz, 32-core processor. Additionally, the corresponding LLL Coulomb energies computed using these three methods are presented in Table~\ref{tab: energies_comparison_N_20}. In the SM~\cite{SM}, we provide a detailed error analysis explaining why the refined method outperforms the naive one. Hereafter, we will only show results obtained using the refined method. 

\begin{table}[htbp!]
    \begin{center}
        \begin{tabular}{ |c|c|c|c|c|c| } 
            \hline
            $\nu$ & $N$ & ED &  MC naive & MC refined & MC full symm\\
            \hline
            $1$ & $20$ & $-0.45422$ &  $-0.44(1)$ & $-0.457(3)$ & $-0.43(2)$\\
            \hline 
        \end{tabular}
    \end{center}
    \caption{The density-corrected~\cite{Morf86} per-particle Coulomb energies for $N{=}20$ bosons in the $\nu{=}1$ Moore-Read state, computed in the spherical geometry by different methods with identical computational resources and the same runtime. The number in the parentheses is the statistical uncertainty coming from the Monte Carlo computations.}
    \label{tab: energies_comparison_N_20}
\end{table}

We now turn our attention to the RR$3$ state. Interestingly, its excitations, Fibonacci anyons, can be used to carry out universal topological quantum computation, in contrast to the Ising anyons of the MR state, which lack the necessary degrees of freedom to achieve universality~\cite{Freedman01, Freedman02}. Importantly, unlike the MR state, where the Pf provides a more efficient alternative to full symmetrization, the RR$3$ wavefunction can only be evaluated by symmetrization. This makes our method particularly well-suited for calculating observables for the RR$3$ state. As in the $k{=}2$ case, we must group permutations corresponding to equivalent matrix elements. However, owing to the presence of three clusters, distinct matrix elements cannot be identified by just exchanges. To overcome this hurdle, we recast the problem in terms of set partitions, which, along with exhaustively enumerating all permutations, help identify unique matrix elements. Each set partition maps to a $3{\times}3$ doubly stochastic integer matrix, which serves as a representative of its equivalence class. These equivalence classes allow us to aggregate terms with identical matrix elements and compute their weights, as detailed in the SM~\cite{SM}. In the $k{=}2$ case, the $2{\times}2$ doubly stochastic integer matrix analysis reduces to using exchanges~\cite{SM}.  

\begin{figure}[htbp!]
\includegraphics[clip,width=0.49\columnwidth]{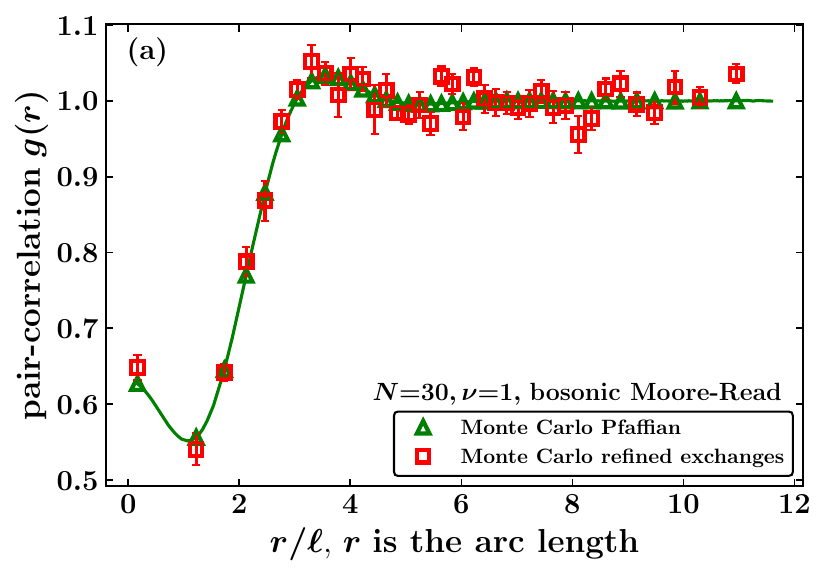} 
\includegraphics[clip,width=0.49\columnwidth]{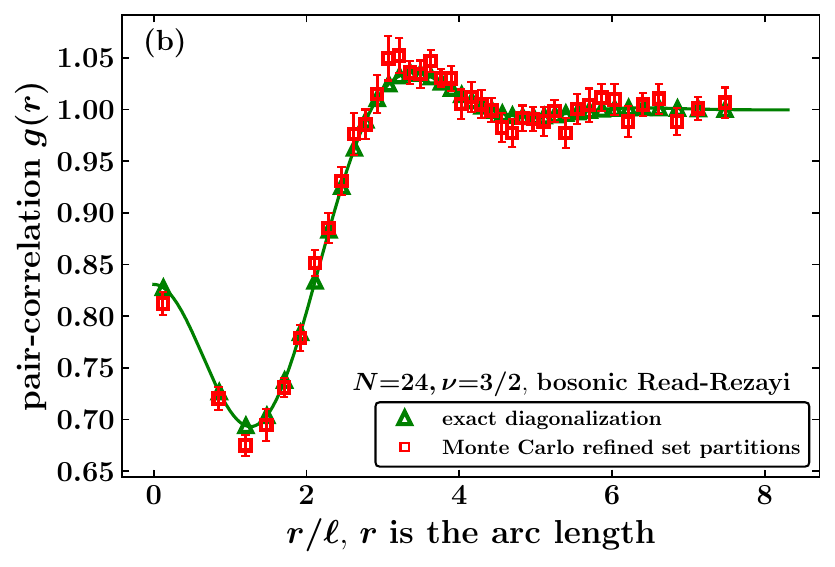}\\ 
\includegraphics[clip,width=0.49\columnwidth]{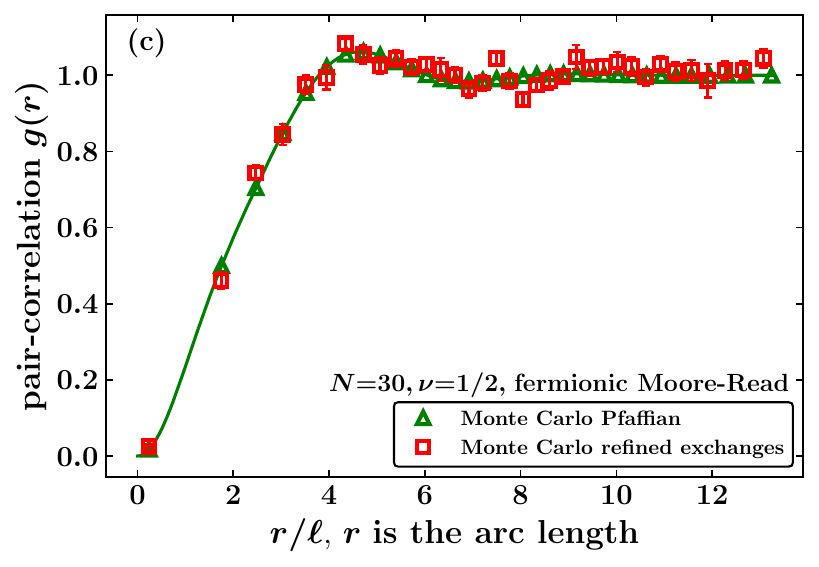}
\includegraphics[clip,width=0.49\columnwidth]{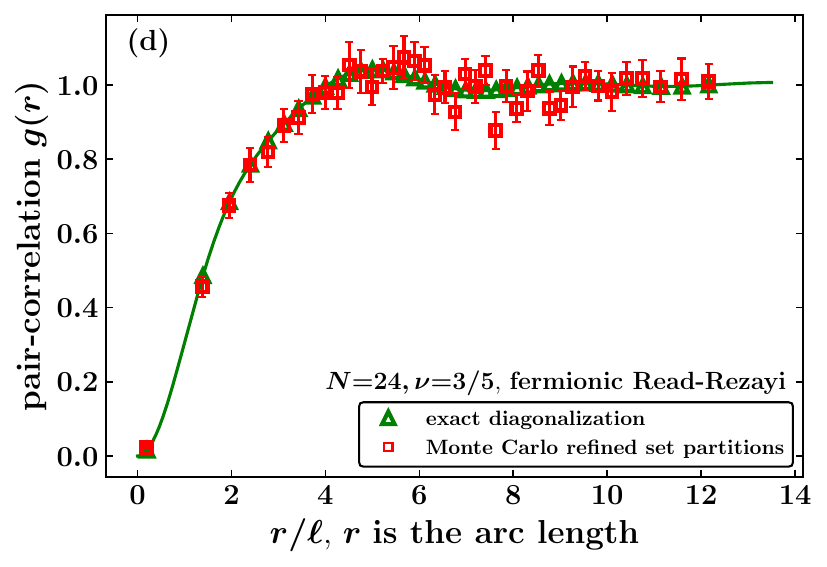}
\caption{Pair-correlation function $g(r)$ for the Moore-Read state of (a) bosons at $\nu{=}1$ and (c) fermions at $\nu{=}1/2$ for $N{=}30$ particles, and the $3$-cluster Read-Rezayi state of (b) bosons at $\nu{=}3/2$ and (d) fermions at $\nu{=}3/5$ for $N{=}24$ particles obtained in the spherical geometry by different methods.}
\label{fig: pair_correlation_bosoncs_fermions_ED_comparison}
\end{figure}

\textbf{\textit{Results}.} 
In Fig.~\ref{fig: pair_correlation_bosoncs_fermions_ED_comparison} and Table~\ref{tab: energies}, we show the pair-correlation function and tabulate the LLL Coulomb energies, respectively, for $N{=}30$ bosons for the $\nu{=}1$ MR and $N{=}24$ bosons for the $\nu{=}3{/}2$ RR$3$ states, obtained in the spherical geometry~\cite{Haldane83}. For each system, we run $20$ independent MC chains. For comparison, MC computations using the Pf and ED results are also shown on the same plot. For $N{=}30$ at $\nu{=}1$, we set $s_{\rm rep}{=}50000$, which results in a total of $\sum_m \mathcal{N}_{(30,m)}{=}1558$ terms, while for $N{=}24$ at $\nu{=}3{/}2$, we set $s_{\rm rep}{=}10^6$, which results in a total of $\sum_{n} \mathcal{N}_{(24,n)}{=}1603$ terms (here $n$ labels the equivalence classes). Our approach is easily extended to the fermionic RR$k$ states since those are related to the bosonic ones by a multiplicative factor of $\Phi_1$. In Fig.~\ref{fig: pair_correlation_bosoncs_fermions_ED_comparison} and Table~\ref{tab: energies}, we also show results for the fermionic MR and RR$3$ states at $\nu{=}1/2$ and $3/5$, respectively, for which we choose the same systems and parameters as the analogous bosonic ones. In all the cases, we find fairly good agreement with the results obtained from the Pf or ED.

\begin{table}[htbp!]
    \begin{center}
        \begin{tabular}{ |c|c|c|c|c| } 
            \hline
            $\nu$ & $N$ & ED or Pf-MC &  MC refined \\
            \hline
            $1$ & $30$ & $-0.45295(1)$ &  $-0.450(2)$ \\ 
            \hline
            $1/2$ & $30$ & $-0.457956(2)$ &  $-0.4586(7)$ \\
             \hline
            $3/2$ & $24$ & $-0.42533$ &  $-0.428(3)$\\
            \hline
            $3/5$ & $24$ & $ -0.4882$ &  $-0.487(1)$\\ 
            \hline 
        \end{tabular}
    \end{center}
    \caption{The density-corrected~\cite{Morf86} per-particle Coulomb energies for $N$ bosons and fermions in the $\nu{=}1$ and $\nu{=}1/2$ Moore-Read and the $\nu{=}3/2$ and $\nu{=}3/5$ $3$-cluster Read-Rezayi states computed in the spherical geometry by different methods. The number in the parentheses is the statistical uncertainty stemming from the Monte Carlo (MC) computations.}
    \label{tab: energies}
\end{table}

\begin{figure}[htbp!]
\includegraphics[clip,width=1\columnwidth]{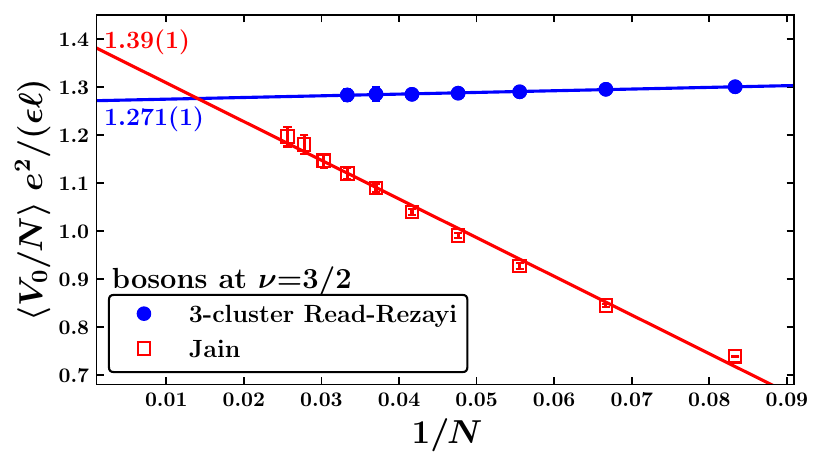} 
\caption{Competition between the bosonic $k{=}3$ Read-Rezayi and Jain states at $\nu{=}3/2$ for the hard-core $V_{0}$ interaction.}
\label{fig: bosonic_RR3_Jain_3_2}
\end{figure}

As an application of our method, we consider the competition between the bosonic RR$3$ and Jain states at $\nu{=}3/2$ with the hard-core delta-function, equivalently, the $V_{0}$ Haldane-pseudopotential~\cite{Haldane83}, interaction. For small systems, the exact $V_{0}$ ground state has a high overlap with the bosonic RR$3$, indicating that RR$3$ could be stabilized for this interaction~\cite{Regnault04}. Here, we go beyond the system sizes accessible to ED and obtain a reliable thermodynamic-limit estimate by evaluating the $V_{0}$ energies of the two states from their density-density correlation, i.e., the static structure factor~\cite{Dora24, Dora25} (the two-body interaction energy of a state is just a weighted sum over the different components of its static structure factor). For the bosonic RR$3$ state, this is done using Monte Carlo with our refined method, while for the Jain state we use standard Monte Carlo~\cite{Kamilla97, Balram17}. The wavefunction of the Jain state is evaluated as the fermionic $3/5$ Jain state~\cite{Jain89} over $\Phi_1$~\cite{Chang05b, Liu20}, since this form is amenable to large-scale numerical evaluation via the Jain-Kamilla projection~\cite{Jain97, Jain97b, Moller05, Davenport12, Gattu24}. The bosonic RR$3$ extrapolates to a lower zero relative angular momentum pair-amplitude than the Jain state, consistent with RR$3$ being stabilized for the $V_0$ interaction~\cite{Regnault04} [see Fig.~\ref{fig: bosonic_RR3_Jain_3_2}].

\textbf{\textit{Discussion}.} 
The ideas we presented can be readily generalized to more clusters. An interesting future direction is to test the competition between the 2/5 Bonderson-Slingerland ~\cite{Bonderson08} and the hole-conjugate of the fermionic RR$3$~\cite{Read99} states for the 12/5 FQHE. Since this competition is very tight~\cite{Bonderson12}, it requires an extremely accurate evaluation of the energies of the RR$3$ state. Extracting topological quantities via this method could be an interesting avenue to explore. For instance, our technique can extend previous studies of braiding statistics~\cite{Nayak96, Tserkovnyak03, Jeon04, Baraban09, Bose24} to the RR$3$ state, to establish that its excitations are Fibonacci anyons~\cite{Read09}. This method can also be used to compute wavefunctions where full (anti)symmetrization is needed, such as at $5/2$ with LL mixing, where the LLL is filled and the second LL is half-filled, or when constructing bipartite or tripartite composite fermion states~\cite{Sreejith11, Sreejith11b, Sreejith13}. Our approach can be further augmented by improving the sampling efficiency of the MC using Hamiltonian or hybrid MC methods~\cite{Duane87, Wang26}. We leave an exploration of this and other directions to future work.

\textbf{\textit{Acknowledgments}.} K.B. thanks Abhimanyu Choudhury for suggesting useful computational ideas. A.C.B. acknowledges useful discussions with Amritanshu Prasad, Anup Dixit, and Arvind Ayyer. The work was made possible by financial support from the Science and Engineering Research Board (SERB) of the Department of Science and Technology (DST) via the Mathematical Research Impact Centric Support (MATRICS) Grant No. MTR/2023/000002 and the Anusandhan National Research Foundation (ANRF) via the Advanced Research Grant No. ANRF/ARG/2025/000562/PS. Computational portions of this research work were conducted using the Nandadevi and Kamet supercomputers maintained and supported by the Institute of Mathematical Sciences' High-Performance Computing Center. Some numerical calculations were performed using the DiagHam package~\cite{diagham}, for which we are grateful to its authors. S.H.S. acknowledges support from EPSRC grant EP/X030881/1. This research was supported in part by the International Centre for Theoretical Sciences (ICTS) for participation in the 10th Indian Statistical Physics Community Meeting (code: ICTS/10thISPCM2025/04).  All data from this work is available at 
github.com/edominzi/symmmetrization\_data.

\bibliography{biblio_fqhe}

\newpage 
\cleardoublepage

\onecolumngrid
\begin{center}
\textbf{\large Supplemental Material for ``Monte Carlo Sampling for Wave Functions Requiring (Anti)Symmetrization"}\\[5pt]

\begin{center}
 {\small Koyena Bose$^{1,2}$, Steven H. Simon$^{3}$ and Ajit C. Balram$^{1,2}$}  
\end{center}

\begin{center}
{\sl \footnotesize
$^{1}$Institute of Mathematical Sciences, CIT Campus, Chennai 600113, India

$^{2}$Homi Bhabha National Institute, Training School Complex, Anushaktinagar, Mumbai 400094, India
    
$^{3}$Rudolf Peierls Centre, Oxford University, OX1 3NP, United Kingdom
}
\end{center}

\begin{quote}
{\small This supplemental material (SM) provides additional details on the results presented in the main text. In Sec.~\ref{sec: set_partitions}, we review the concept of set partitions, which provides a systematic way to group permutations that yield identical matrix elements. We then detail how to identify the unique equivalence classes of integrals and derive their corresponding weights for both the Moore-Read and $3$-clustered Read-Rezayi states. In Sec.~\ref{sec: error}, we provide an error analysis of the naive, refined, and full symmetrization methods. We derive a formula that quantifies how the error grows with the number of additional terms included in the refined method and show that it agrees well with numerically computed error. We also analyze how the time required to achieve a target accuracy scales with system size.}
\end{quote}
\end{center}

\vspace*{0.4cm}

\setcounter{equation}{0}
\setcounter{figure}{0}
\setcounter{table}{0}
\setcounter{page}{1}
\setcounter{section}{0}
\makeatletter
\renewcommand{\theequation}{S\arabic{equation}}
\renewcommand{\thefigure}{S\arabic{figure}}
\renewcommand{\thesection}{S\Roman{section}}
\renewcommand{\thepage}{\arabic{page}}
\renewcommand{\thetable}{S\arabic{table}}

\section{Set partitions}
\label{sec: set_partitions}
We define a base representation, i.e., the identity permutation, for $N$ particles divided into $k$ groups as $G{=}G_1 {\cup} G_2 {\cup} {\cdots} {\cup} G_k$, with $G_i{=}\{(i{-}1)N/k{+}1, (i{-}1)N/k{+}2,{\cdots},iN/k\} $ for $i{=}1,2,{\cdots},k$. All distinct exchanges of particle indices between these groups then generate the complete set of permutations required for (anti)symmetrization.

\subsection{$k{=}2$ cluster case}
For $k{=}2$, we generate these exchanges using the combinatorial identity
\begin{equation}
\label{eq: N_k_2_exchange_bimomial_identity1}
    N! = \left[\left( \frac{N}{2} \right)!\right]^{2} \sum_{m{=}0}^{N/2} \binom{N/2}{m}^{2}.
\end{equation}
The left-hand side of Eq.~\eqref{eq: N_k_2_exchange_bimomial_identity1} corresponds to all possible permutations of $N$ particles, while the right-hand side explicitly counts permutations generated by exchanges between the two clusters, each of which has $N/2$ elements. Specifically, for the $m^{\rm th}$ exchange, we choose $m$ particles from each of the two clusters and exchange them, with $m{=}0, 1, {\cdots}, N/2$. The appropriate combinatorial factor for $m$ exchanges is
\begin{equation}
\label{eq: combinatorial_factor_k_2}
C_{(N,m)}{=}\binom{N/2}{m}{\times}\binom{N/2}{m} \equiv C_{(N,N/2-m)}.    
\end{equation}
Accounting for all possible orderings within each cluster yields a factor of $(N/2)!$ per cluster, which results in Eq.~\eqref{eq: N_k_2_exchange_bimomial_identity1}. Importantly, as far as expectation values of permutation-invariant operators, referred to as the integrals, are concerned, exchanging $m$ indices produces the same result as exchanging $N/2{-}m$ indices, as this is equivalent to swapping the two clusters. Thus, the minimal set of exchanges is generated by taking $m{=}0,1,2,{\cdots},\lfloor N/4 \rfloor$, where $\lfloor x \rfloor{=}{\rm floor}(x)$ is the greatest integer${\leq}x$, while the $m'{=}\lfloor N/4 \rfloor{+}1,{\cdots}, N/2$ exchanges are accounted for by adding their contribution to $m{=}N/2{-}m'$. This results in a reduction of Eq.~\eqref{eq: N_k_2_exchange_bimomial_identity1} to
\begin{equation}
\label{eq: N_k_2_exchange_bimomial_identity3}
    \frac{N!}{2!\left[\left( \frac{N}{2} \right)!\right]^{2}} = 
    \begin{cases}
    \sum_{m=0}^{\lfloor N/4 \rfloor} C_{(N,m)} , & \text{if $N$ is not a multiple of $4$} \\
    \frac{C_{(N, N/4)}}{2} + \sum_{m=0}^{N/4{-}1} C_{(N,m)} ,  & \text{if $N$ is a multiple of $4$}
    \end{cases},
\end{equation}
where the left-hand side counts the minimal number of permutations over which the (anti)symmetrization has to be carried out. A representative for the $m^{\rm th}$ exchange is obtained by simply swapping the first $m$ elements in the base set $G_1$ with those of $G_2$. 

\subsection{$k{=}3$ cluster case}
To group permutations that result in the same integral, we classify them into equivalence classes as follows. The base representation is defined as $G{=}G_1G_2G_3$, where $G_1{=}\{1,{\cdots},N/3\}$, $G_2{=}\{N/3{+}1,{\cdots},2N/3\}$, and $G_3{=}\{2N/3{+}1,{\cdots},N\}$, while any generic permutation is given by $G'{=}G_1'G_2'G_3'$. To determine which equivalence class a given permutation belongs to, we construct a $3{\times}3$ matrix $M$ corresponding to the given permutation, such that $M_{ij}{=}|G_i {\cap} G^{'}_j|$, where $|H|$ is the cardinality/size of the set $H$. Every such matrix $M$ represents a multi-dimensional integral $I(M)$, which for the bosonic 3-clustered Read-Rezayi state is given by:
\begin{equation}
     I(M)=\int \prod_{i=1}^{N}d^2 z_{i}~f(\{z_{i}\})~\Phi^{2}_1(\{z(G_1)\})\Phi^{2}_1(\{z(G_2)\})\Phi^{2}_1(\{z(G_3)\})~ 
     \Phi^{2}_1(\{z(G^{'}_1)\})\Phi^{2}_1(\{z(G^{'}_2)\})\Phi^{2}_1(\{z(G^{'}_3)\}),
\end{equation}
where $f({z_{i}})$ is a permutation invariant function and $\Phi^{2}_1(\{z(G_1)\}){=}\prod_{1{\leq}i{<}j{\leq}N/3}(z_{i}{-}z_{j})$, and so on. These matrices $M$ are known as doubly stochastic integer matrices since they have non-negative integer entries and their row and column sums are exactly $N/3$. The number of unique $3{\times}3$ doubly stochastic integer matrices, each representing an equivalence class here, as a function of $N$ has been enumerated previously~\cite{OEIS_A052282, Beck2015}. Matrices related by row or column permutations are considered equivalent here since these operations correspond to swapping $G_{i}{\leftrightarrow}G_{j}$ or $G'_{i}{\leftrightarrow}G'_{j}$, respectively. This invariance is justified as the integrals themselves are unchanged under such permutations, as each $G_{i}$ or $G^{'}_j$ enters in the same functional form of $\Phi^{2}_1$. Since each $3{\times}3$ matrix has $3!{\times}3!{=}36$ forms owing to row and column permutations, there are essentially $36$ representations of a matrix all belonging to the same equivalence class. However, not all of the $36$ matrices generated by row and column permutations, though belonging to the same equivalence class, are distinct in terms of their matrix elements. For example, in the case of $N{=}6$, there are $3$ equivalence classes. Let us look at one of them, corresponding to the identity permutation, given by
\begin{equation}
\label{eq: M_for_N_6_identity}
    M_{(N=6,n=1)}= \begin{bmatrix}
     2 & 0 & 0 \\
     0 & 2 & 0\\
     0 & 0 & 2
    \end{bmatrix}.
\end{equation}
Here, the subscript $N{=}6$ stands for the number of particles and the label $n{=}1$ denotes an equivalence class. Applying the same permutation to both the rows and columns of $M_{(6,1)}$ given in Eq.~\eqref{eq: M_for_N_6_identity} preserves its diagonal structure. Specifically, the following permutations leave the form of the matrix invariant: 
$$\{ (\mathbb{I}, \mathbb{I}), (01,01), (02,02), (12,12), (012,012), (021,021)\},$$
where $(abc,xyz)$ denotes a simultaneous row and column permutation with rows $a{\rightarrow}b{\rightarrow}c{\rightarrow}a$ and columns $x{\rightarrow}y{\rightarrow}z{\rightarrow}x$, respectively. Therefore, this matrix has $6$ symmetry operations which leave it unchanged, forming its Automorphism group ${\rm Aut}(M_{(6,1)})$, with cardinality $|{\rm Aut}(M_{(6,1)})|{=}6$. Similarly, let us consider another example of an equivalence class for $N{=}9$, described by the matrix
\begin{equation}
    M_{(9,2)}= \begin{bmatrix}
     1 & 1 & 1 \\
     1 & 1 & 1\\
     1 & 1 & 1
    \end{bmatrix},
\end{equation}
wherein all row and column permutations leave the matrix invariant. Therefore, this automorphism group has cardinality $|{\rm Aut}(M_{(9,2)})|{=}36$. From these considerations, we can compute the weights associated with each equivalence class using Burnside's lemma~\cite{Huppert2025} as:
\begin{equation}
\label{eq: N_k_3_weight}
    \mathcal{C}^{(k{=}3)}_{(N,n)}= \frac{3!}{|{\rm Aut}(M_{(N, n)})|} \times \frac{[(N/3)!]^3}{\prod_{i,j} (M_{(N, n)})_{ij}!}.
\end{equation}
We utilize the fact that each group $G_i$ contains $N/3$ particles, which are distributed in clusters of $(M_{(N, n)})_{ij}$, where $j{=}1,2,3$. The number of ways to carry out this distribution is $(N/3)!/\prod_j (M_{(N, n)})_{ij}!$, and doing so across all three groups results in $[(N/3)!]^3/\prod_{i,j}(M_{(N, n)})_{ij}!$ ways. Since each $3{\times}3$ matrix has $3!{\times}3!{=}36$ forms owing to row and column permutations, the weight corresponding to each form must be multiplied by $36$ to obtain the total weight. However, not all row and column permutations necessarily produce distinct matrices as discussed above.  Thus, to obtain the correct weight, we divide by the size $|{\rm Aut}(M)|$, which represents the number of permutation operations that leave the matrix invariant. Finally, since the three groups $G_1, G_2, G_3$ can be permuted among each other (all these permutations result in the same wave function), we divide the final result by $3!$ to avoid over-counting. In Table~\ref{tab: N_15_matrices_weights}, we have listed out all the equivalence classes along with the cardinality of the corresponding automorphism groups and their weights for $N{=}15$. 

\begin{table}[tbhp!]
\centering
\begin{tabular}{|c|c|c|c|}
\hline
serial number, $n$ & matrix $M_{(15,n)}$ representing an equivalence class  & $|{\rm Aut}(M_{(15, n)})|$ & weight, $\mathcal{C}^{(3)}_{(15,n)}$ \\
\hline
1 & $\begin{bmatrix}0 & 0 & 5 \\ 0 & 5 & 0 \\ 5 & 0 & 0\end{bmatrix}$ & 6 & 1 \\
\hline
2 & $\begin{bmatrix}0 & 0 & 5 \\ 1 & 4 & 0 \\ 4 & 1 & 0\end{bmatrix}$ & 2 & 75 \\
\hline
3 & $\begin{bmatrix}0 & 0 & 5 \\ 2 & 3 & 0 \\ 3 & 2 & 0\end{bmatrix}$ & 2 & 300 \\
\hline
4 & $\begin{bmatrix}0 & 1 & 4 \\ 1 & 3 & 1 \\ 4 & 1 & 0\end{bmatrix}$ & 2 & 1500 \\
\hline
5 & $\begin{bmatrix}0 & 1 & 4 \\ 1 & 4 & 0 \\ 4 & 0 & 1\end{bmatrix}$ & 3 & 250 \\
\hline
6 & $\begin{bmatrix}0 & 1 & 4 \\ 2 & 2 & 1 \\ 3 & 2 & 0\end{bmatrix}$ & 1 & 9000 \\
\hline
7 & $\begin{bmatrix}0 & 1 & 4 \\ 2 & 3 & 0 \\ 3 & 1 & 1\end{bmatrix}$ & 1 & 6000 \\
\hline
8 & $\begin{bmatrix}0 & 2 & 3 \\ 2 & 1 & 2 \\ 3 & 2 & 0\end{bmatrix}$ & 2 & 9000 \\
\hline
9 & $\begin{bmatrix}0 & 2 & 3 \\ 2 & 2 & 1 \\ 3 & 1 & 1\end{bmatrix}$ & 1 & 36000 \\
\hline
10 & $\begin{bmatrix}0 & 2 & 3 \\ 2 & 3 & 0 \\ 3 & 0 & 2\end{bmatrix}$ & 3 & 2000 \\
\hline
11 & $\begin{bmatrix}1 & 1 & 3 \\ 1 & 3 & 1 \\ 3 & 1 & 1\end{bmatrix}$ & 6 & 8000 \\
\hline
12 & $\begin{bmatrix}1 & 1 & 3 \\ 2 & 2 & 1 \\ 2 & 2 & 1\end{bmatrix}$ & 4 & 27000 \\
\hline
13 & $\begin{bmatrix}1 & 2 & 2 \\ 2 & 1 & 2 \\ 2 & 2 & 1\end{bmatrix}$ & 6 & 27000 \\
\hline
\end{tabular}
\caption{The $13$ equivalence classes for the $3{-}$cluster Read-Rezayi state for $N{=}15$ particles. Each equivalence class is represented by a doubly stochastic $3{\times}3$ integer matrix, and its automorphism group cardinality (see text) and the corresponding weights are also given [see Eq.~\eqref{eq: N_k_3_weight}].}
\label{tab: N_15_matrices_weights}
\end{table}


\subsection{General $k{-}$cluster case}
For general $k$, equivalence classes corresponding to a system size of $N$ can be determined using $k{\times}k$ doubly stochastic integer matrices with row and column sums $N/k$, equivalent up to row and column permutations~\cite{OEIS_A333733}. The weights corresponding to each equivalence class can similarly be determined using combinatorial arguments as above, which results in
\begin{equation}
\label{eq: N_k_weight}
    \mathcal{C}^{(k)}_{(N,n)}= \frac{k!}{|{\rm Aut}(M_{(N, n)})|} \times \frac{[(N/k)!]^k}{\prod_{i,j} (M_{(N, n)})_{ij}!}.
\end{equation}
The weight given in Eq.~\eqref{eq: N_k_weight} satisfies the combinatorial identity $\sum_{n}\mathcal{C}^{(k)}_{(N,n)}{=}N!/[k! \left(\left(N/k\right)!\right)^k]$, which precisely counts the minimal number of permutations over which the (anti)symmetrization needs to be done. Furthermore, given a doubly stochastic $k{\times}k$ matrix, it is straightforward to generate representative permutations for the equivalence class that it stands for.

Interestingly, the treatment used in the $k{=}2$ case to determine the number of unique exchanges is the same as the number of equivalence classes of $2{\times}2$ doubly stochastic integer matrices. In this case, the matrix is determined by a single parameter, the number of exchanges $m$, since the remaining entries are fixed by the row and column sums. Furthermore, the weight corresponding to $m$ exchanges given in Eq.~\eqref{eq: N_k_2_exchange_bimomial_identity3} match the weights of the equivalence classes given by 
\begin{equation}
\label{eq: N_k_2_exchange_doubly_stochastic_identity1}
    \mathcal{C}^{(2)}_{(N,m)}= \frac{2!}{|{\rm Aut}(M_{(N, m)})|} \times \frac{[(N/2)!]^2}{\prod_{i,j} (M_{(N, m)})_{ij}!},
\end{equation}
i.e., for $m{\neq}N/4$, $\mathcal{C}^{(2)}_{(N,m)}{=}(C_{(N, m)}{+}C_{(N, N/2{-}m)})/2{\equiv}C_{(N, m)}$, and $\mathcal{C}^{(2)}_{(N,N/4)}{=}C_{(N, N/4)}/2$ for $m{=}N/4$, where $C_{(N, m)}$ is defined in Eq.~\eqref{eq: combinatorial_factor_k_2}. Thus, for the $k{=}2$ case, the clubbing of permutations solely based on exchanges between clusters matches exactly with the clubbing of equivalent integrals via $2{\times}2$ doubly stochastic integer matrices.

\section{Error analysis}
\label{sec: error}
In this section, we provide an error analysis of our method for the $k{=}2$ case. We quantify how the error scales with the number of representations used per equivalence class. Firstly, let us consider the method where full symmetrization (full-symm) is done. By this we mean, we look at the expectation value $\langle \hat{O} \rangle_{\Psi^{\rm MR}_{1}}{=} \frac{1}{N!}  \langle \Psi^{{\rm H}(2,2,0)}_{1} | \hat{O}  \sum_m  \sum_{t=0}^{C_{(N,m)}} |\Psi^{{\rm H}(2,2,0),m_t}_{1} \rangle$), where the sum over $m$ and $t$ put together run over the $\binom{N}{N/2}/2$ permutations, i.e., the full set of permutations is used on the ket side. Throughout, we always use $\Psi^{{\rm H}(2,2,0)}_{1}$ as the sampling function. The statistical error in the Monte Carlo (MC) computation of $\langle \hat{O} \rangle_{\Psi^{\rm MR}_{1}}$ is
\begin{equation}
\label{eq: epsilon_full_symm}
    \epsilon_{\rm full-symm}= \epsilon/\sqrt{M},
\end{equation}
where $M$ is the total number of MC steps, and the total error $\epsilon$ is
\begin{equation}
\label{eq: epsilon_full_symm_independence}
    \epsilon=\sum_{i=1}^n\epsilon_i=\sum_{i=1}^{n} \sum_{j=1}^{C_{(N,i)}} \epsilon_{i,j},
\end{equation}
where $n$ is the total number of equivalence classes, $C_{(N,i)}$ is the total weight of the $i^{\rm th}$ class for a system of size $N$, and  $\epsilon_{i,j}{=}\sqrt{(\langle I_i \rangle {-} I_{i,j})^2}$, i.e., $\epsilon_{i,j}$ is the error due to the $I_{i,j}$ integral coming from the $j^{\rm th}$ representative of the $i^{\rm th}$ class. To arrive at Eq.~\eqref{eq: epsilon_full_symm_independence}, we used the independence of each term in the sum. Thus, following Eq.~\eqref{eq: epsilon_full_symm}, the total error is
\begin{equation}
    \epsilon_{\rm full-symm}= \sum_{i=1}^{n} \frac{\epsilon_i}{\sqrt{M}}. 
\end{equation}

Now, when only a portion of the representative permutations, referred to as representations from here on in, is used, as in the naive and refined methods discussed in the main text, the total error is
\begin{equation}
    \epsilon_{\rm reduced}=\frac{1}{\sqrt{M}}\sum_{i=1}^{n} \frac{C_{(N,i)}}{\mathcal{N}_{(N,i)}}\sum_{j=1}^{\mathcal{N}_{(N,i)}} \epsilon_{i,j}
\end{equation}
We express $\epsilon_{i,j}{=}\langle \epsilon_{i} \rangle{+}\Delta_{i,j}$, where $\langle \epsilon_{i} \rangle{=}\epsilon_i/C_{(N,i)}$ is the average error of the $i^{\rm th}$ class and $\Delta_{i,j}$ is the deviation of each term $\epsilon_{i,j}$ from its average. The quantity $\mathcal{N}_{(N, i)}$ represents the number of terms retained from the $i^{\rm th}$ equivalence class, and the factor $C_{(N, i)}/\mathcal{N}_{(N, i)}$ is the adjusted weight factor to account for the missing terms in each equivalence class. When all the representations of an equivalence class are included, then $\sum_{j} \Delta_{i,j}{=}0$, as in the full-symm case. However, when only a few representatives of an equivalence class are used, the choice of the representations creates a bias, leading to an error
\begin{eqnarray}
    \epsilon_{\rm reduced}&=&\frac{1}{\sqrt{M}}\sum_{i=1}^{n} \frac{C_{(N,i)}}{\mathcal{N}_{(N,i)}}\sum_{j=1}^{\mathcal{N}_{(N,i)}} \big ( \langle \epsilon_{i} \rangle+\Delta_{i,j} \big), \nonumber \\ 
    &=& \frac{1}{\sqrt{M}}\sum_{i=1}^{n} \frac{C_{(N,i)}}{\mathcal{N}_{(N,i)}} \mathcal{N}_{(N,i)} \frac{\epsilon_i}{C_{(N,i)}} + \frac{1}{\sqrt{M}}\sum_{i=1}^{n} \sum_{j=1}^{\mathcal{N}_{(N,i)}}  \frac{C_{(N,i)}}{\mathcal{N}_{(N,i)}} \Delta_{i,j} \nonumber \\
    &\approx&\epsilon_{\rm full-symm} + \frac{1}{\sqrt{M}} \sum_{i=1}^{n}  \frac{C_{(N,i)}}{\mathcal{N}_{(N,i)}}  \sum_{j=1}^{\mathcal{N}_{(N,i)}} \frac{ \Delta  } {\sum_{k=1}^{n} \mathcal{N}_{(N,k)}}.
\end{eqnarray}
where, $\Delta{=}\sum_{i{=}1}^{n}\sum_{j{=}1}^{\mathcal{N}_{(N, i)}}\Delta_{i,j}$ and $\langle \Delta\rangle{=}\Delta/\sum_{i=1}^{n} \mathcal{N}_{(N,i)}$ is the average over all clases. We replace the individual $\Delta_{i,j}$ by its average over all classes, which gives us the bias, $\mathcal{B}_{\rm reduced}$, as
\begin{eqnarray}
\label{eq: bias_reduced}
   \mathcal{B}_{\rm reduced}=\epsilon_{\rm reduced}-\epsilon_{\rm full-symm}&\approx&\frac{1}{\sqrt{M}}\frac{\sum_{i=1}^{n} C_{(N,i)}}{\sum_{k=1}^n\mathcal{N}_{(N,k)}} \Delta \nonumber \\
    \implies\mathcal{B_{\rm reduced}} &\approx& \frac{1}{\sqrt{M}}\frac{\mathcal{S}}{\mathcal{E}} \Delta,
\end{eqnarray}
where $\mathcal{S}$ is the total number of terms in full-symm given by $N!^2/(2!(N/2)!^2)$ and $\mathcal{E}$ is the total terms kept in the naive or refined method. 

\begin{figure}[htbp!]
\includegraphics[clip,width=0.49\columnwidth]{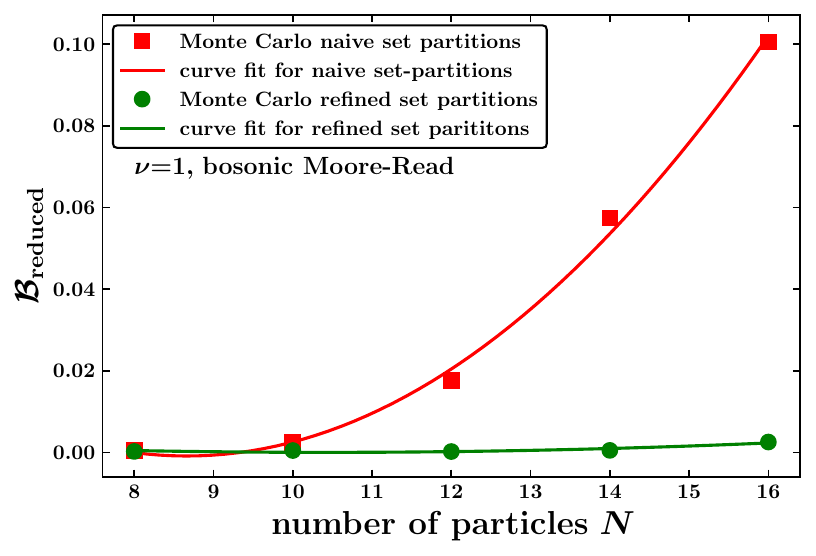}
\includegraphics[clip,width=0.49\columnwidth]{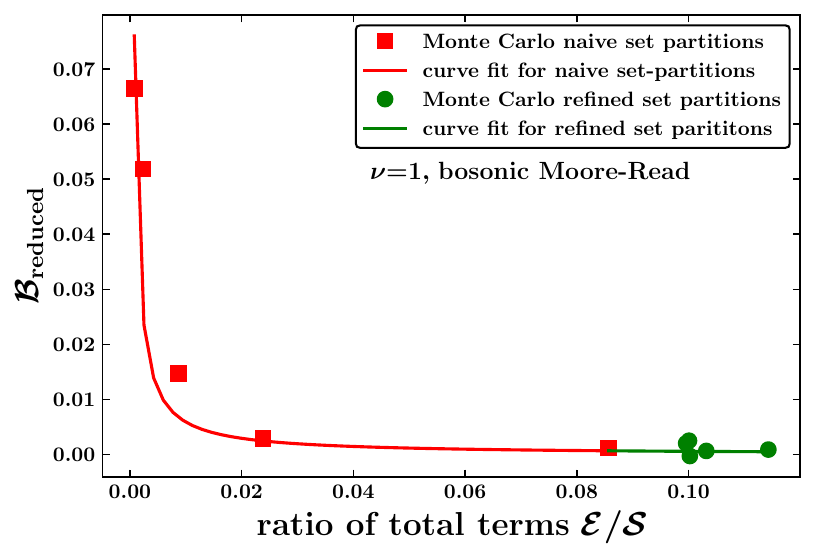}
\caption{The bias $\mathcal{B}_{\rm reduced}$ [see Eq.~\eqref{eq: bias_reduced}] as a function of number of particles $N$ [left panel], as well as $\mathcal{E}/\mathcal{S}$ [right panel], for system sizes $N{=}\{ 8,10,12,14,16 \}$ for the bosonic Moore-Read state at $\nu{=}1$ for the naive and refined methods with the same number of Monte Carlo iterations. }
\label{fig: bias_comparison}
\end{figure}

To test the validity of Eq.~\eqref{eq: bias_reduced}, we numerically compute the statistical error for both the naive and the refined methods across different system sizes $N{=}\{ 8,10,12,14,16 \}$ by calculating their lowest Landau level Coulomb energy in the spherical geometry. For the naive method, the total number of terms is $\lfloor N/4 \rfloor{+}1$, as described in the main text. For the refined method, we include additional terms for each system such that $\mathcal{E}/\mathcal{S}{\approx}0.1$. For each system, for every term included in the refinement, we run the same number of MC steps to ensure a fair comparison. In Fig.~\ref{fig: bias_comparison}, we show the plot of the numerically computed bias for the naive and the refined methods as a function of $N$. The bias in the naive method rapidly grows with increasing $N$, while that in the refined method remains almost constant since the ratio $\mathcal{E}/\mathcal{S}$ is kept nearly fixed to $0.1$ for all $N$. Furthermore, motivated by Eq.~\eqref{eq: bias_reduced}, we plot the bias as a function of $\mathcal{E}/\mathcal{S}$, and fit it to the hyperbolic $b{\times}\mathcal{E}/\mathcal{S}$ form  [for clarity we use $\mathcal{E}/\mathcal{S}$, instead of its inverse $\mathcal{S}/\mathcal{E}$ that appears in Eq.~\eqref{eq: bias_reduced}], where $b$ is a fitting parameter. We find a good fit to the numerically obtained data, showing that the bias is inversely proportional to $\mathcal{E}$, as one would anticipate, i.e., the bias is reduced as more terms are kept in the refinement. For both the naive and refined methods, the numerically obtained data fall onto a single curve when fitted with a single parameter $b$ to the $\mathcal{E}/\mathcal{S}$ form, thereby highlighting the dependence of the bias-induced error solely on this ratio. From this result, we understand that the statistical error in the reduced form, i.e., when only a portion of the representative permutations is kept, is comprised of two parts: (1) the MC statistical error and (2) the error due to bias. 

A reasonable and practically relevant question to ask is how many iterations $M'$ does the refined method need to reach a given statistical error $\epsilon$ when $\mathcal{E}$ terms are included, compared to the number of MC steps $M$ needed under full symmetrization to obtain the same error $\epsilon$. Thus, we set $\epsilon_{\rm reduced}{=}\epsilon/\sqrt{M}$ and $\epsilon_{\rm full-symm}{=}\epsilon/\sqrt{M'}$ in Eq.~\eqref{eq: bias_reduced} to get
\begin{eqnarray}
\label{eq: MC_steps}
     \frac{\epsilon}{\sqrt{M}} &\approx& \frac{\epsilon}{\sqrt{M'}}+\frac{ \Delta}{\sqrt{M'}}\frac{\mathcal{S}}{\mathcal{E}}\nonumber \\
     \implies M' &\approx& \bigg (1+ \frac{\mathcal{S}}{\mathcal{E}}\frac{\Delta}{\epsilon}\bigg)^2 M.
\end{eqnarray}
In Fig. 1 of the main text, we show that for a fixed set of computation resources [including cores and CPU time in a parallelized code], the refined method works better than the naive one. The key reason is that the ratio $\mathcal{E}/\mathcal{S}$ is significantly higher in the refined method compared to the naive one. As an example, let us take $N{=}30$ for $k{=}2$, which has $\mathcal{S}{=}77558760$ and $\mathcal{E}_{\rm naive}{=}8$, so $\mathcal{S}/\mathcal{E}_{\rm naive}{=}9694845$ and consider $\Delta/\epsilon{=}1$ (Note that we do not know the precise value of $\Delta/\epsilon$, but we expect it to be an order $1$ number.). The number of MC steps required by the naive method to reach the desired convergence is $M'{\approx}10^{14}M$ [see Eq.~\eqref{eq: MC_steps}], i.e., about $10^{14}$ times higher than what would be required for full symmetrization. For the refined case, $\mathcal{E}_{\rm refined}{=}1558$, therefore, $\mathcal{S}/\mathcal{E_{\rm refined}}{=}49780.97$. Assuming $\Delta/\epsilon$ remains the same, we will only need $2.5{\times}10^9$ times the full symmetrization MC steps. Thus, although the refined method lowers the MC step count by a factor $4{\times}10^4$ compared to the naive method, each step is about $1558/8{\approx}200$ times slower due to the requirement of computation of the additional terms. While the inclusion of extra terms reduces the number of MC steps required to reach a given accuracy, this advantage is somewhat offset by the increased cost per step. However, if the code is parallelized to compute the $\mathcal{E}_{\rm refined}$ terms (which will not speed up the naive method as it has very few representations), such as using $64$ cores, the final result can still be approximately $10^4$ times faster than the naive method. Hence, the choice of $\mathcal{E}_{\rm refined}$ (or $s_{\rm rep}$) must be made carefully, as inflating the number of extra terms per equivalence class, $\mathcal{N}_{(N,m)}$, for larger values of $N$, can quickly cause the runtime to grow from hours to days or even months. Therefore, depending on the available computational resources and wall-time limits, one can adjust $\mathcal{E}_{\rm refined}$ (or $s_{\rm rep}$) to achieve the best trade-off between error reduction and speed. We note that Graphical Processing Units (GPUs) could be ideally suited to carry out these computations, as their extensive and efficient parallel processing power can be leveraged to evaluate the many refined terms simultaneously.

\begin{figure}[htbp!]
\includegraphics[clip,width=0.49\columnwidth]{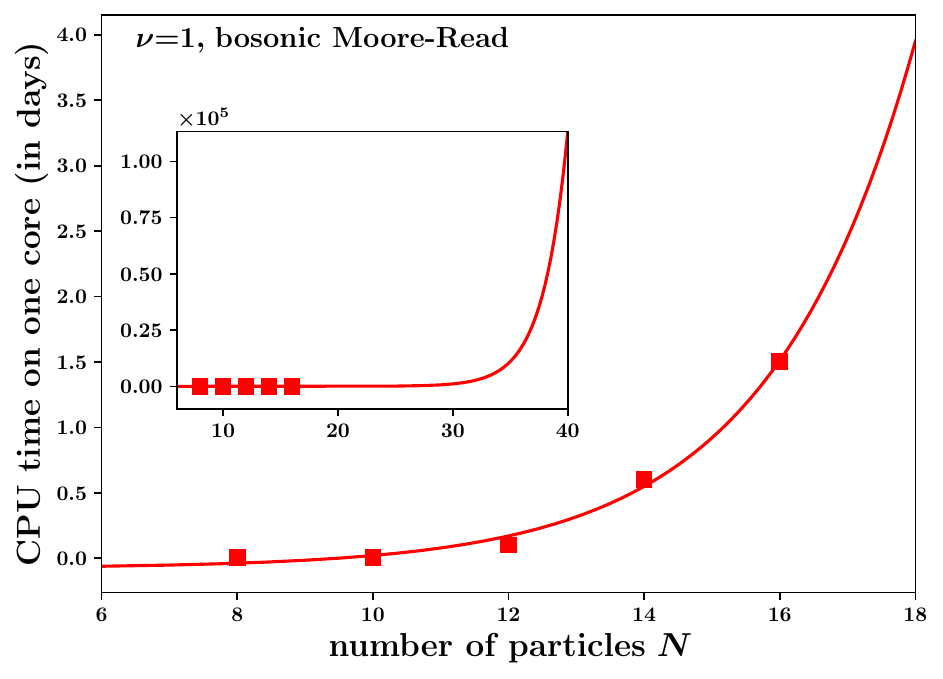}
\caption{The CPU time per core required to converge the static structure factor computed in spherical geometry to within a standard deviation of $0.01$ from the exact result for the $\nu{=}1$ bosonic Moore-Read state for various systems of $N$ bosons on the sphere using the refined method with the number of extra terms $\mathcal{E}{=}\lceil 0.4e^{0.3912N} \rceil$. The red squares represent the numerically obtained results for $N{=}\{8,10,12,14,16\}$, while the red line represents the best-fit exponential, ${-}108.83{+}1.33e^{0.46N}$. The inset shows the curve fitting extrapolated to a system size of up to $40$ particles to give a rough estimate of its convergence time (it has the same $x$-axis and $y$-axis labels as the main plot). }
\label{fig: error_comparison}
\end{figure}

As this trade-off between accuracy and speed is central, it is useful to estimate the time required to achieve a desired level of convergence across different system sizes using the refined method. To this end, we consider a specific choice of the number of extra terms as $\mathcal{E}{=}\lceil ae^{0.3912N}\rceil$, with $a{=}0.4$ and $b{=}0.3912$, chosen to make best use of the computational resources available to us, and where $\lceil x \rceil{=}{\rm ceil}(x)$ is the least integer${\geq}x$. In Fig.~\ref{fig: error_comparison}, we show the time required for system sizes $N{=}\{8,10,12,14,16\}$ to reach an error $\sqrt{\sum_{L}|S_{\rm exact}(L){-}S_{\rm refined}(L)|^{2}}$ of $0.01$ in the static structure factor computation, where $L$ is the total orbital angular momentum on the sphere~\cite{Haldane83}, using a single core for the refined method (denoted by the red squares). The data nicely fits a function of the form $a{+}be^{cN}$, with $a{=}{-}108.83,~b{=}1.33$ and $c{=}0.46$ (denoted by the red line). Since $e^c{\approx}1.58{<}2$, the effective scaling exponent is slightly smaller than what is typically observed in exact diagonalization (ED), allowing us to access a few more system sizes compared to ED. For $N{=}40$, the estimated time to reach an accuracy of $0.01$ in the structure factor is about six months when running $10$ MC chains with 64 cores each (obtained by dividing the single-core CPU time by $10{\times}64$). While the speed can be improved by allocating more cores per MC chain, the exponential growth of CPU time with system size still implies that simulations beyond $N{=}40$ are expected to become rapidly intractable, even with increased parallel computational resources. 

\end{document}